\documentclass[prd,reprint,nofootinbib,showpacs,superscriptaddress]{revtex4-1}
\usepackage{graphicx} 
\usepackage{hyperref}
\usepackage{amsfonts}
\usepackage{amsmath,amssymb}
\usepackage{bm} 
\usepackage{color}
\usepackage{epstopdf} 
\usepackage{epsfig}
\usepackage{subfig} 

{\rm }

\def\be{\begin{equation}}
 \def\ee{\end{equation}}
 \def\bea{\begin{eqnarray}}
 \def\eea{\end{eqnarray}}
 \def\bes{\begin{eqnarray}}
 \def\ees{\end{eqnarray}}
 \def\bi{\begin{itemize}}
 \def\ei{\end{itemize}} 

\def\2{\frac{1}{2}}
\def\4{\frac{1}{4}}

\begin{document}

\title{Free-space reconfigurable quantum key distribution network}

\author{Bing Qi}
\email{qib1@ornl.gov}
\affiliation{Oak Ridge National Laboratory, Oak Ridge, Tennessee, USA}
\affiliation{University of Tennessee, Knoxville, Tennessee, USA}

\author{Hoi-Kwong Lo}
\affiliation{University of Toronto, Toronto, Ontario, Canada}

\author{Charles Ci Wen Lim}
\affiliation{Oak Ridge National Laboratory, Oak Ridge, Tennessee, USA}

\author{George Siopsis}
\affiliation{University of Tennessee, Knoxville, Tennessee, USA}

\author{Eric A. Chitambar}
\affiliation{Southern Illinois University, Carbondale, Illinois, USA}

\author{Raphael Pooser}
\affiliation{Oak Ridge National Laboratory, Oak Ridge, Tennessee, USA}
\affiliation{University of Tennessee, Knoxville, Tennessee, USA}

\author{Philip G. Evans}
\affiliation{Oak Ridge National Laboratory, Oak Ridge, Tennessee, USA}

\author{Warren Grice}
\affiliation{Oak Ridge National Laboratory, Oak Ridge, Tennessee, USA}


\begin{abstract}
We propose a free-space reconfigurable quantum key distribution (QKD) network to secure communication among mobile users. Depends on the trustworthiness of the network relay, the users can implement either the highly secure measurement-device-independent QKD, or the highly efficient decoy state BB84 QKD. Based on the same quantum infrastructure, we also propose a loss tolerant quantum position verification scheme, which could allow the QKD users to initiate the QKD process without relying on pre-shared key. 
\end{abstract}

\maketitle

\section{Introduction}

Like it or not, today's wireless communication allows you to be connected anytime anywhere. Comparing with communication through optical fibers, a free-space communication system is more susceptible to eavesdropping due to the openness of the communication channel, thus there is a more urgent need to enhance its security. Quantum key distribution (QKD) \cite{LCT:natphoton14} allows two remote users (Alice and Bob) to generate cryptographic keys with proven security through an insecure channel. A mobile QKD network could bring unprecedented level of security to wireless users. 

Despite its enormous potential, mobile QKD network has however received limited attention. Most of today's quantum key distribution (QKD) experiments are conducted through optical fiber links, in light of the availability of worldwide optical fiber network. However, to apply QKD in a mobile communication network, free-space transmission seems the only viable option.

Extensive research has been conducted in free-space QKD, aiming at building up a global QKD network by using satellites as trusted relays \cite{NMR:natphoton13,WYL:natphoton13}. In those studies, the main goal is to establish a long-distance QKD link between a stationary laboratory at the earth and a satellite in space moving along a well-defined orbit at a relatively small angular speed. In contrast, a mobile QKD network may only cover a relatively small geographic area but is highly dynamic in nature: both the locations of QKD users and the corresponding accessible network relays are time dependent. In this case, the trustworthiness of the network relay may be questionable. Furthermore, as the QKD users may frequently move among different networks, a secure authentication scheme without relying on pre-shared key is highly desired.

In the paper, we propose a free-space reconfigurable QKD network based on the recently discovered measurement-device-independent (MDI) QKD protocol \cite{LCQ:prl12}. The MDI-QKD is an ideal building block for multi-user QKD network, since the most expensive and vulnerable measurement device can be placed in an untrusted network relay and be shared by many QKD users. Another advantage of our proposal is its reconfigurability: depending on the trustworthiness of the network relay, the QKD users can implement either the highly secure MDI-QKD or the highly efficient decoy-state BB84 QKD \cite{Hwang:prl03,LMC:prl05,Wang:prl05}. This feature is especially appealing when the measurement device is implemented with low-efficiency detectors. Furthermore, based on the same quantum infrastructure, we also propose a loss-tolerant quantum position verification scheme secure against ``local'' quantum adversaries, which could allow a legitimate party to use its geographical location as its only credential to establish an authenticated channel.

\section{MEASUREMENT-DEVICE-INDEPENDENT (MDI) QKD}
\label{sec:2}

Idealized QKD protocols have been proved to be unconditionally secure against adversaries with unlimited computing power and technological capabilities. However, real-life implementations of QKD rarely conform to the assumptions in idealized models used in the security proofs. Indeed, by exploiting security loopholes in practical realizations, especially imperfections in the detectors, several attacks have been successfully launched against commercial QKD systems \cite{ZFQ:pra08,LWW:natphoton10,XQL:njp10}. Compared with an optical fiber link, a free-space optical link is more vulnerable, in the sense that it is much more easily accessible via physical means. This suggests that a free-space QKD system could be more vulnerable to side-channel attacks. For example, in the same spirit as the time-shift attack \cite{QFL:qic07}, an eavesdropper can implement a ``space-shift'' attack by simply manipulating the spatial mode of the quantum signal, as first proposed in \cite{FTQ:qic09} and demonstrated recently in \cite{SCB:pra15}.

In a conventional QKD protocol (see Fig.1a), Alice prepares quantum signals and Bob performs measurement. In this case, the errors in quantum state preparation can be well controlled and quantified, since this can be done within the Alice's well-protected laboratory without Eve's interference. On the other hand, the quantum states received by the measurement device are highly unpredictable, since Eve can replace the original quantum states with anything at her will. This makes the measurement device the most vulnerable part is the whole QKD system.

\begin{figure}[htb]
\centerline{\includegraphics[width=8cm]{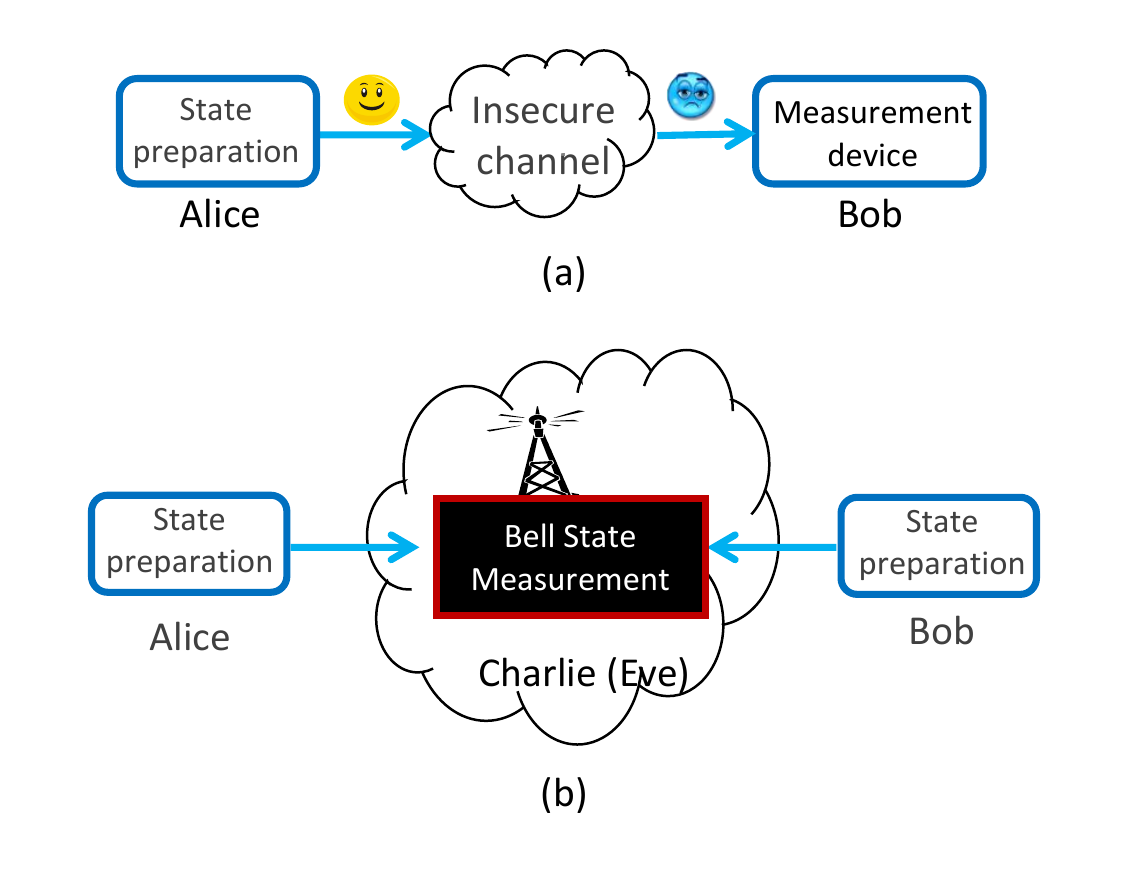}} \caption{(a) Conventional QKD; (b) Measurement-device-independent (MDI) QKD}
\end{figure}

The motivation behind MDI-QKD is to develop a QKD protocol which is automatically immune to all detector side-channel attacks. In fact, the measurement device in the MDI-QKD can be treated as a ``black box'' which could even be manufactured and operated by the eavesdropper. In this scheme (see Fig.1b), both Alice and Bob prepare BB84 \cite{BB84} quantum states and send them to an untrusted third party, Charlie, who performs Bell state measurement and publicly announces the measurement results. Given Charlie's measurement results, Alice and Bob can further establish a secure key. The security of MDI-QKD is based on the idea of time-reversed EPR QKD \cite{BHM:pra96,Inamori:alg02}: a successful Bell state measurement at Charlie projects Alice and Bob's photons into an EPR state. If Charlie executes the protocol honestly, he cannot gain any information of the secure key. On the other hand, any attempts by Charlie to gain information of the secure key will destroy the quantum correlation expected from an EPR state and thus can be detected by Alice and Bob. 

MDI-QKD is an ideal building block of mobile QKD networks, since the most expensive measurement device can be placed inside a network relay while each QKD user only needs a low-cost transmitter. In fact, the recent progresses in chip-scale QKD \cite{SEG:arxiv15} suggest that a QKD transmitter could be fitted into compact mobile devices such as smart phones. This could allow a QKD user to use a simple handhold transmitter to generate quantum key for cryptographic purposes. Furthermore, since the security does not rely on the trustworthiness of the network relays, the QKD users can optimize key distribution route based on their locations without compromising the security. This protocol is highly practical and can be implemented with off-the-shelf components. Up-to-date, MDI-QKD has been demonstrated over 200 km telecom fiber link \cite{TYC:jstqe14} and a quantum channel with 60 dB loss \cite{VLC:arxiv15}. 

MDI-QKD completely removes any potential security loopholes on the detector side, so the remaining question is how to quantify imperfections at the source and take them into account in the security proof. A promising solution is the loss-tolerant protocol \cite{TCK:pra14} proposed by Tamaki et al., which makes QKD tolerable to channel loss in the presence of source flaws. Recently, the loss-tolerant protocol has been applied into a MDI-QKD experimental demonstration \cite{TWB:arxiv15}. Combined with existing countermeasures against Trojan-horse attack \cite{LCW:arxiv15}, this approach could lead to practical side-channel-free QKD.

Although the above MDI-QKD experiments were conducted through optical fiber links, there are no fundamental roadblocks in its free-space application. Especially, polarization encoding scheme, the most favorable choice for free-space QKD, has been demonstrated in a complete MDI-QKD experiment \cite{TLX:prl14}. Nonetheless, a number of research opportunities/challenges do exist. For example, owing to the atmospheric turbulence, the time synchronization and mode matching of two separate free-space channels with respect to a common relay, Charlie, might be non-trivial and may lead to a substantial reduction in the transmittance and an increase in quantum bit error rate. Procedures for time synchronization and mode matching may need to be developed to overcome these challenges.

\section{FREE-SPACE RECONFIGURABLE QKD NETWORK}
\label{sec:3}

MDI-QKD can significantly improve the security of practical QKD.  The price to pay is the relatively low key generation rate when implemented with low-efficiency detectors. In conventional QKD, such as the decoy state BB84 protocol, the secure key rate is approximately proportional to the detector efficiency. On the other hand, in MDI-QKD, secure keys are generated from two-fold coincidence detection events, the secure key rate is approximately proportional to the square of detector efficiency. So the MDI-QKD suffers more from the low detection efficiency of conventional single photon detector. We remark that high-efficiency superconducting nanowire single photon detector (SNSPD) is commercially available \cite{SNSPD}. If such high-efficient detectors are used, as discussed in \cite{XCQ:arxiv15}, the key rate of MDI-QKD can be high enough for metropolitan network application.

Another challenge in free-space MDI-QKD, which has been briefly discussed in previous section, is how to match the temporal and spatial modes of photons from different users. To achieve a high interference visibility in Bell state measurement (thus a high secure key rate in MDI-QKD), photons from two QKD users should be highly indistinguishable. Given the atmospheric turbulence acts independently in the two quantum channels, it could be very challenging to match the arrival times and spatial modes of photons propagated through two independent free-space channels precisely.

\begin{figure}[htb]
\centerline{\includegraphics[width=8cm]{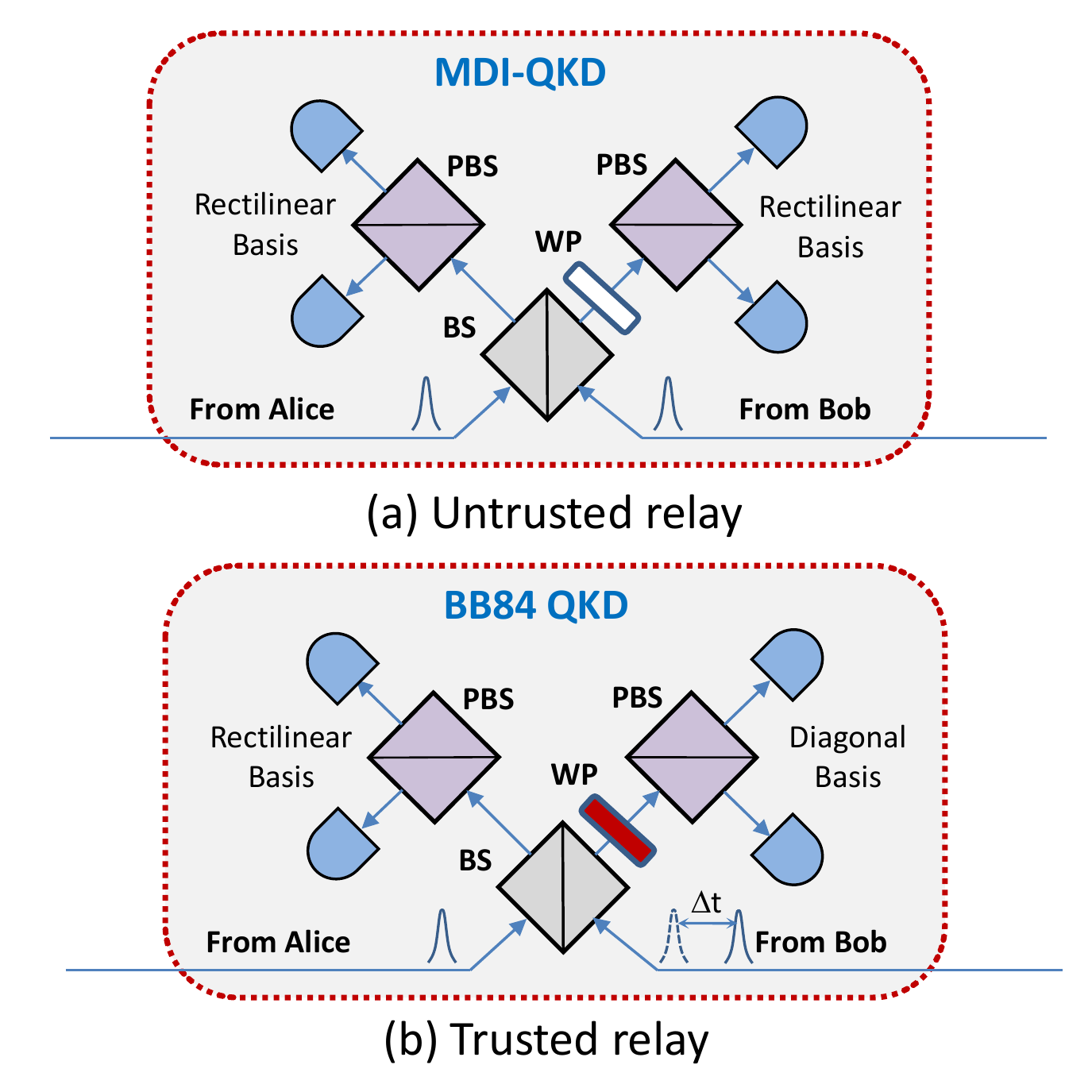}} \caption{(a) MDI-QKD implemented with an untrusted network relay; (b) BB84 QKD implmented with a trusted network relay. BS-Beamsplitter; PBS-Polarizing beamsplitter; WP-Waveplate. }
\end{figure}

To alleviate the above problems, we propose a reconfigurable QKD which can easily switch between the highly secure MDI-QKD and the highly efficiency decoy state BB84 QKD: when the QKD users move into an untrusted network, they can implement MDI-QKD protocol using the measurement device in an untrusted network relay, as shown in Fig.2a. On the other hand, whenever the network relay can be fully trusted, the QKD users can use the trusted relay to implement decoy state BB84 protocol: each QKD user generates a secure key with the relay independently. After that, the network relay can establish a secure channel with Bob by using the key shared with Bob to implement one-time-pad and convey the key shared with Alice to Bob. As shown in Fig.2, the above two protocols can be implemented using the same hardware design. To switch from the MDI-QKD to the decoy state BB84, the network relay can simply delay the photons from one user to make sure photons from different users hit the beam splitter at different times. He can also rotate a wave plate to change the polarization measurement basis, as shown in Fig.2b. By providing the users with the flexibility of implementing multiple protocols on the same platform, the above QKD network design could be fit into many practical applications.

\section{LOSS TOLERANT QUANTUM POSITION VERIFICATION}
\label{sec:3}

According to special relativity, information cannot be transmitted at a speed faster than the speed of light in vacuum. This fundamental constrain can open the door to novel applications in free space line of sight optical communications. One potential application is the position based cryptography (PBC), where a legitimate user can use its geographical location as the only credential to implement various cryptographic protocols \cite{BCF:siam14}. A position based authentication scheme may allow QKD users to establish a classical authenticated channel without using pre-shared key. Such a protocol could be very useful in a mobile QKD network where the users may frequently switch between different networks.

The basic task of PBC is position verification, where a prover $P$ tries to convince a group of trusted verifiers that he/she is at the claimed location. Intuitively, position verification can be implemented as follows: each verifier sends one piece of information to $P$ and the verifiers coordinate their transmission time to make sure that $P$ receives all the information simultaneously. The prover $P$ performs certain operations/calculations using the information from the verifiers and reports the result back. If all the verifiers receive the correct answer at the time consistent with the claimed location, the location of $P$ is accepted as authenticated. 

Unfortunately, it was shown in \cite{CGM:lnc09} that all the classical position verification protocols (where each verifier can only send classical information to $P$) are vulnerable to attacks from a coalition of adversaries possessing only classical communication channels: each adversary can intercept the classical information sent by the nearest verifier and forward a perfect copy of the intercepted information to each of his/her partners. After collecting all the information, each adversary can perform the operations/calculations expected from $P$ and report the result to the nearest verifier.

Quantum position verification (QPV) protocols, where the verifiers are allowed to send quantum information to $P$, have been developed with the hope to achieve information-theoretic security \cite{BCF:siam14, KBM:patent06, KMS:pra11, Kent:pra11, Malaney:pra10, LL:pra11}. On one hand, it has been shown that a QPV protocol can be unconditionally secure under attacks of a coalition of adversaries limited to local operations and classical communications; on the other hand, it has also been shown that any PBQC protocols are breakable if the adversaries have unlimited quantum resources, such as entanglement \cite{BCF:siam14}. It is an active research topic to study the security of QPV protocols under attacks of adversaries with limited quantum resources \cite{BK:njp11, TFK:njp13, RG:arxiv15, CL:arxiv15}, a more relevant scenario in practice.

Inspired by MDI-QKD, we propose a QPV protocol based on Bell state measurement \cite{Note1}. This protocol can be implemented using the same setup for MDI-QKD and thus can be easily integrated into the proposed free-space QKD network. This protocol is also loss-tolerant, an important advantage in practical applications \cite{QS:pra15}. For simplicity, we consider the one-dimensional case with two verifiers, $V_0$ and $V_1$, and a prover P is in between. The basic procedures of the proposed PBQC protocol are as follows:

\begin{enumerate}

\item Through a private channel, the verifiers $V_0$ and $V_1$ agree on random bits $x_0, x_1, \theta \in \{ 0,1 \}$. $V_0$ prepares a qubit in the state,
\be |\psi_0 \rangle = H^{\theta} |x_0\rangle, \ee
where $|0\rangle$ and $|1\rangle$ are computational basis states, and $H$ is the Hadamard matrix. Similarly, $V_1$ prepares a qubit in the state,
\be |\psi_1 \rangle = H^{\theta} |x_1\rangle. \ee

Note, in our scheme, $V_0$ and $V_1$ choose encoding basis randomly and collaboratively.

\item $V_0$ ($V_1$) sends $|\psi_0\rangle$ ($|\psi_1\rangle$) to $P$ through a free-space quantum channel. $V_0$ and $V_1$ coordinate their transmission times to make sure that $|\psi_0\rangle$ and $|\psi_1\rangle$ arrive at $P$ at the same time.

\item As soon as $|\psi_0\rangle$ and $|\psi_1\rangle$ arrive, $P$ performs a Bell-state measurement. If $P$ detects a Bell-state successfully, he broadcasts the measurement result to $V_0$ and $V_1$ immediately through free-space classical communication channels. Otherwise, he reports no detection.

\item If the verifiers receive the measurement results at a time inconsistent with the position of $P$, the protocol fails and will be terminated.

\item Through an authenticated classical channel, $V_0$ and $V_1$ compare the reported measurement results received by them. If they receive different results, the protocol fails and will be terminated.

As noted in \cite{LL:pra11}, noisy operation in the honest case do not produce inconsistent results between different verifiers. This is because the classical communication channels used by $P$ to broadcast the measurement results are virtually noiseless. By comparing the results received by $V_0$ and $V_1$, we can further limit the adversaries' power \cite{QS:pra15}. 

\item $V_0$, $V_1$ and $P$ repeat the above procedures many times. They calculate the error rate $E_R$, which is defined as the probability that the reported measurement result is inconsistent with the quantum states transmitted. If $E_R$ is below a predetermined value, the location of $P$ is accepted as authenticated.
\end{enumerate}

Like previous QPV protocols, the proposed QPV protocol based on Bell state measurement is insecure if the adversaries share entanglement or they possess quantum communication channels (which allows them to exchange quantum signals, such as EPR state). In appendix A, we provide an intuitive security analysis under attacks of a coalition of adversaries limited to local operations and classical communications (A more detailed security proof under the same assumptions will appear in forthcoming publication). By introducing decoy state, we expect that the above protocol can be implemented with practical weak coherent sources and realistic detectors without sacrificing its performance, as the case of MDI-QKD. To extend this protocol into 2 or 3 dimensional spaces, we could introduce more verifiers to detect the classical measurement results broadcast by the prover, as shown in Fig.3: two verifiers send BB84 photons to $P$, who performs Bell measurement and broadcast the measurement results through free-space classical communication channels. If all the three verifiers receive the same measurement result at times consistent with the position of $P$ and the measurement result is consistent with the quantum states transmitted, the location of $P$ is accepted as authenticated.

\begin{figure}[htb]
\centerline{\includegraphics[width=8cm]{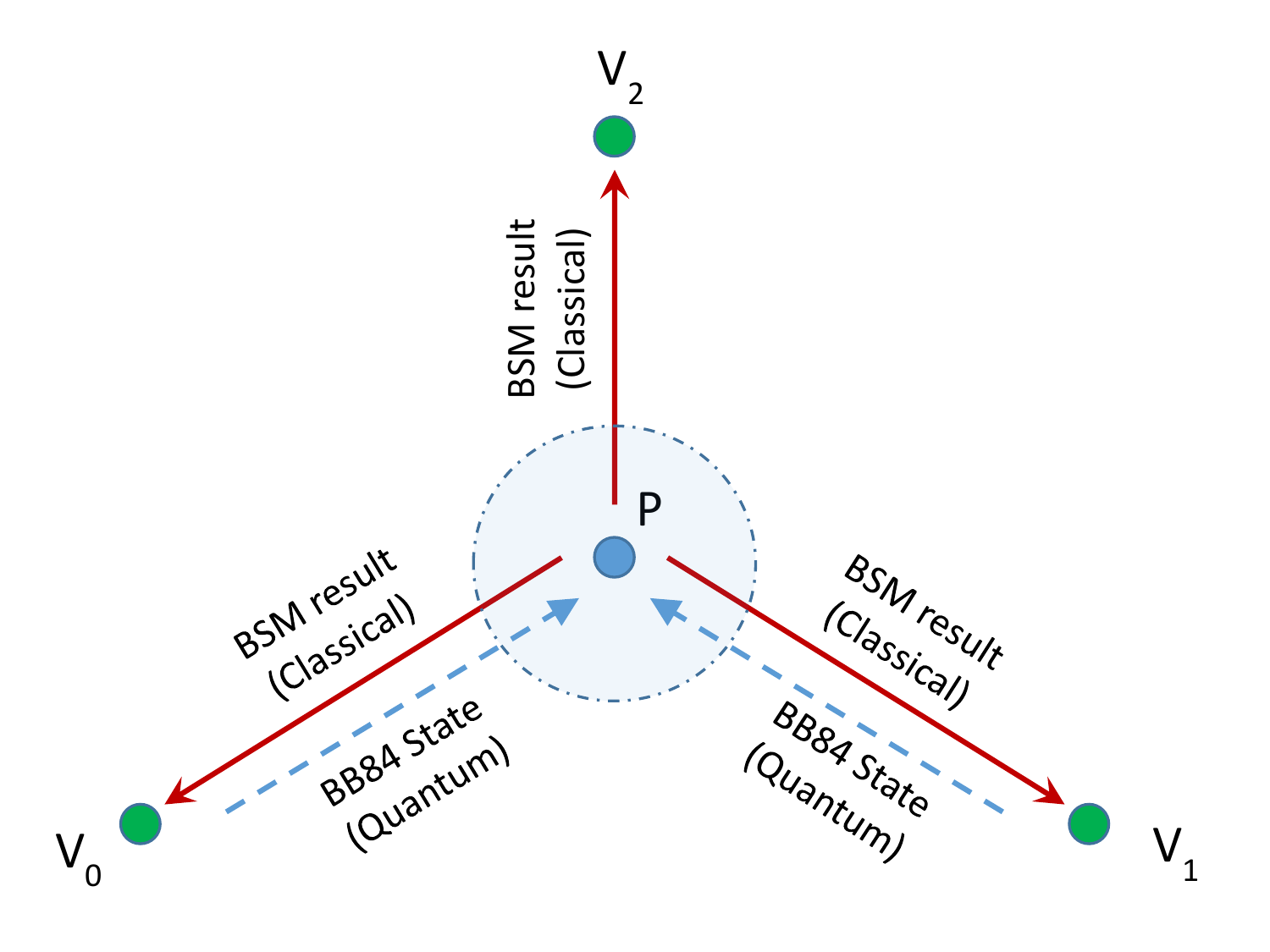}} \caption{Qaunutm positon verification based on Bell state measurement}
\end{figure}

\section{OUTLOOK}
\label{sec:dis}

In light of the availability of worldwide optical fiber network, most of QKD experiments are conducted through optical fiber links. In practice, there is also an urgent need to secure communication in mobile networks. Here we propose a free-space reconfigurable mobile QKD network based on loss-tolerant MDI-QKD protocol. We expect such a solution will find wide applications in the future.

\acknowledgments{
Part of this work was performed at Oak Ridge National Laboratory, operated by UT-Battelle for the U.S. Department of Energy under Contract No. DE-AC05-00OR22725.}

\appendix

\section{Towards a loss tolerant QPV scheme assuming LOCC adversaries}

For simplicity, we consider the one-dimensional case with two verifiers, $V_0$ and $V_1$, and a prover $P$ in between. There are two adversaries, $E_0$ (between $V_0$ and $P$) and $E_1$ (between $V_1$ and $P$). 

Like previous QPV protocols, the proposed QPV protocol based on Bell state measurement is insecure if the adversaries share entanglement or they possess quantum communication channels (which allows them to exchange quantum signals, such as EPR state), as highlighted by the attack below:

Suppose for simplicity that only the Bell state $\vert \Psi^+ \rangle$ will be reported by the honest $P$. If $E_0$ and $E_1$ pre-share an EPR pair in the state $\vert \Psi^+ \rangle$, they can launch the following attack:

(1)As soon as $E_0$ ($E_1$) intercepts the photon from $V_0$ ($V_1$), she performs a Bell measurement on it and the photon of the EPR pair in her possession.

(2)$E_0$ ($E_1$) sends her (classical) Bell measurement results to $E_1$ ($E_0$).

(3) If both of them detect $\vert \Psi^+ \rangle$, $E_0$ ($E_1$) reports the measurement results to $V_0$ ($V_1$). Otherwise, they claim no detection. 

Evidently, the above attack will not introduce any errors and thus cannot be detected when the expected detection rate from $P$ is low.

Here we consider a weaker security model in which the adversaries are restricted to local quantum operations and classical communication (LOCC). This choice of security model allows us to establish security bounds that are loss-independent, however they are not valid against more general attacks. Nevertheless, we believe that the LOCC security model is still of practical interest, especially in environments whereby reliable distribution of entanglement is challenging.

We first introduce an entanglement-based PBQC protocol which is equivalent to the prepare-and-measure protocol presented in the main text. In the entanglement-based protocol, both $V_0$ and $V_1$ hold perfect Einstein-Podolsky-Rosen (EPR) photon pairs. Through a private channel, $V_0$ and $V_1$ agree on a randomly chosen basis (either the computational basis or the diagonal basis). Each of them measures one photon of the EPR at hand in the chosen basis, records the measurement result, and sends the other photon to $P$. The rest steps are the same as the prepare-and-measure protocol presented in the main text. From the adversaries' point of view, they cannot distinguish the entanglement-based protocol from the prepare-and-measure PBQC protocol.

Since the measurements performed by $V_0$ and $V_1$ commute with the Bell-state measurement by $P$ (or the measurements preformed by the adversaries), we can switch their order and delay the measurements at $V_0$ and $V_1$ till the end of the protocol. In this picture, a general attack under the LOCC model is as follows: $E_0$ ($E_1$) intercepts half of $V_0$ ($V_1$)'s EPR pair, performs an optimal measurement, and forwards the (classical) measurement results to $E_1$ ($E_0$). Based on the measurement results acquired by both $E_0$ and $E_1$, they follow a pre-determined strategy to either report a specific Bell state or claim no detection. Their goal is to minimize the error rate.

Since we are allowed to delay the measurements performed by $V_0$ and $V_1$, right after $E_0$ ($E_1$)'s measurement, the photon possessed by $V_0$ and $V_1$ will be projected into certain quantum states about which the adversaries may have partial or complete information. Here we assume that by post-selecting the favorable cases based on their measurement results, the adversaries can prepare the photon at $V_0$ ($V_1$)'s hand into any \textit{pure} state at their will. This is the most-favorable assumption for the adversaries, which may or may not be achievable.

Without the loss of generality, we assume that $E_0$ and $E_1$ will report a Bell state $|\Psi^{-}\rangle$ when they prepare the joint state at $V_0$ and $V_1$ as

\be |\varphi \rangle = (\alpha_0 |0\rangle + \beta_0 |1\rangle)_{V_0}\otimes (\alpha_1 |0\rangle + \beta_1 |1\rangle)_{V_1}, \ee
where $\vert\alpha_0\vert^2+\vert\beta_0\vert^2=\vert\alpha_1\vert^2+\vert\beta_1\vert^2=1$.

Suppose $V_0$ and $V_1$ measure photons at their hands in either the computational or diagonal basis with the same probability. It is straightforward to show the error rate in the computational basis (which is the probability that $V_0$ and $V_1$ have the same measurement result) is given by
\be E_{R_1} = \vert\alpha_0\alpha_1\vert^2+\vert\beta_0\beta_1\vert^2. \ee

In the diagonal basis, the joint state at $V_0$ and $V_1$ is given by 
\begin{eqnarray} 
 |\varphi \rangle & = &  \dfrac{1}{2} [(\alpha_0 + \beta_0) |+\rangle + (\alpha_0 - \beta_0) |-\rangle]_{V_0} \nonumber \\ 
 & & \otimes [(\alpha_1 + \beta_1) |+\rangle + (\alpha_1 - \beta_1) |-\rangle]_{V_1}.
\end{eqnarray}

So the error rate in the diagonal basis (which is the probability that $V_0$ and $V_1$ have the same measurement result) is given by
\be E_{R_2} =\dfrac{1}{4} \lbrace \vert(\alpha_0+\beta_0)(\alpha_1+\beta_1)\vert^2+\vert(\alpha_0-\beta_0)(\alpha_1-\beta_1)\vert^2 \rbrace. \ee

After some algebraic manipulation, the average error rate can be expressed as
\begin{eqnarray} 
E_R & = &  \dfrac{E_{R_1}+E_{R_2}}{2} \nonumber \\ 
&& = \dfrac{1+ \vert(\alpha_0\alpha_1+\beta_0\beta_1)\vert^2+\vert(\alpha_0\alpha_1^*+\beta_0\beta_1^*)\vert^2}{4}.
\end{eqnarray}

Obviously, the minimum average error rate introduce by the above attack is $25\%$. This result is remarkable since as long as the intrinsic error rate of a PBQC system is below $25\%$, the location of $P$ can be authenticated regardless the overall loss. The suggests that our protocol is loss-tolerant.

\bibliography{ICSOS_qi}
\bibliographystyle{apsrev}

\end{document}